\documentclass[aip,apl,reprint,groupedaddress,preprintnumber,showpacs]{revtex4-1}
\usepackage{amsmath}
\usepackage{amssymb}
\usepackage{graphicx}
\usepackage{amssymb}
\usepackage{graphics}
\usepackage{graphicx}
\usepackage{epsfig}
\usepackage{CJK}
\usepackage{color}
\usepackage{multirow}
\usepackage{float}

\begin{document}

\begin{CJK*}{GBK}{}
\title{High Fermi velocities and small cyclotron masses in LaAlGe}
\author{Zhixiang Hu,$^{1,2}$ Qianheng Du,$^{1,2}$  Yu Liu,$^{1,\P}$ D. Graf,$^{3}$ and C. Petrovic$^{1,2,\ddag}$}
\affiliation{$^{1}$Condensed Matter Physics and Materials Science Department, Brookhaven National Laboratory, Upton, New York 11973, USA\\
$^{2}$Department of Materials Science and Chemical Engineering, Stony Brook University, Stony Brook, New York 11790, USA\\
$^{3}$National High Magnetic Field Laboratory, Florida State University, Tallahassee, Florida 32306-4005, USA}
\date{\today}

\begin{abstract}
We report quantum oscillation measurements of LaAlGe, a Lorentz-violating type-II Weyl semimetal with tilted Weyl cones. Very small quasiparticle masses and very high Fermi velocities were detected at the Fermi surface. Whereas three main frequencies have been observed, angular dependence of two Fermi surface sheets indicates possible two-dimensional (2D) character despite the absence of the 2D structural features such as van der Waals bonds. Such conducting states may offer a good platform for low-dimensional polarized spin current in magnetic RAlGe (R = Ce, Pr) materials.
\end{abstract}

\maketitle
\end{CJK*}

Unlike normal semimetals, topological Dirac semimetals have surface states which are induced by topology of the bulk.\cite{Wehling,Hasan} The states can be described by linear energy dispersion near Dirac points in the Brillouin zone with very small effective mass and high carrier mobility. In crystals additional symmetry breaking turns Dirac points into pairs of Weyl nodes with distinct chirality that are separated in momentum space. Surface states connecting two different Weyl nodes form Fermi arcs whereas Weyl nodes represent monopoles of Berry curvature, giving rise to unusual electronic transport phenomena.\cite{WeylH,WanX,LvB,ZhangCL,YanB}

This is of interest for a broad range of applications.\cite{RajamathiC,ChanC2} The type-II Weyl semimetals with tilted Weyl cones in three-dimensional (3D) bulk crystals attract considerable attention due to tilted dispersion; the Fermi surface features electron and hole pockets that touch at the Weyl node.\cite{SoluyanovA,BergholtzE,TrescherM,AliM,Bomantara} The dispersion violates Lorentz symmetry and gives rise to tunable properties that could be used in room-temperature optoelectronics, valley filtering and photodetector fabrication.\cite{ChiS,ChanC,SieE,Yesilyurt,ZhangM,ZhangK,ShiS,Nguyen} Of particular interest are optical aplications similar to graphene; topological semimetal crystals host Dirac surface plasmon polariton and allow for propagation of electromagnetic modes in a waveguide geometry.\cite{Thakur,Kotov,Mola,LiHeng}

LaAlGe features type-II Weyl nodes.\cite{ChangG,XuSY,UdagawaM} Thin films of LaAlGe show metallic resistivity, small residual resistivity ratio RRR = $\rho$(300 K)/$\rho$ (5 K) = 1.17, relatively high residual resistivity of about 87 $\mu$$\Omega$ cm, single band electronic transport and 7$\times$10$^{21}$cm$^{-2}$ carrier concentration at 5 K.\cite{Bhattarai} In particular, contribution of Weyl points at the Fermi surface, dimensionality and characteristics of such states are important. Here we perform quantum oscillation studies of LaAlGe. Our results indicate that rather small effective masses and high Fermi velocities of tilted type-II Weyl cones are dominant in electronic transport. We also present evidence for quasi-2D Fermi surface states despite the absence of van der Waals gap or other two-dimensional atomic units in the crystal structure. LaAlGe Fermi surface can be further tuned in magnetic members of RAlGe family (R = Ce, Pr) since they are ferromagnets where Zeeman field could be treated as a coupling without altering the low-energy band structure of LaAlGe.\cite{ChangG}

Single crystals of LaAlGe were grown by high temperature self-flux method.\cite{Fisk} La,Ge and Al were mixed in a ratio of 1:1:20 in an alumina crucible and sealed in a vacuum inside a quartz tube. The ampoule was heated to 1175 $^{\circ}$C for 55 hours and kept there for two hours. Shiny crystals of about 2 mm $\times$ 2 mm $\times$ 0.5 mm were decanted from liquid at 700 $^{\circ}$C after one week of cooling. Excess residual flux was cleaned by polishing before measurements. Powder X-ray diffraction (XRD) was performed on crushed crystal at room temperature by using Cu-K$_\alpha$ ($\lambda$ = 0.15418 nm) radiation in a Rigaku Miniflex powder diffractometer. Resistivity was measured by conventional four-wire method in a Quantum Design PPMS-9. Small cuboid specimen were taken to National High Magnetic Field Laboratory (NHMFL) for measurements of temperature and angular dependent de Haas-van Alphen (dHvA) oscillations, where the field ranged from 0 to 35 Tesla. The crystals were mounted onto miniature Seiko piezoresistive cantilevers which were installed on a rotating platform. The field direction can be changed continuously between parallel ($\theta$ = 0$^{\circ}$) and perpendicular ($\theta$ = 90$^{\circ}$) to the $c$-axis of the crystal.

\begin{figure}[!htb]
\centerline{\includegraphics[scale=0.35]{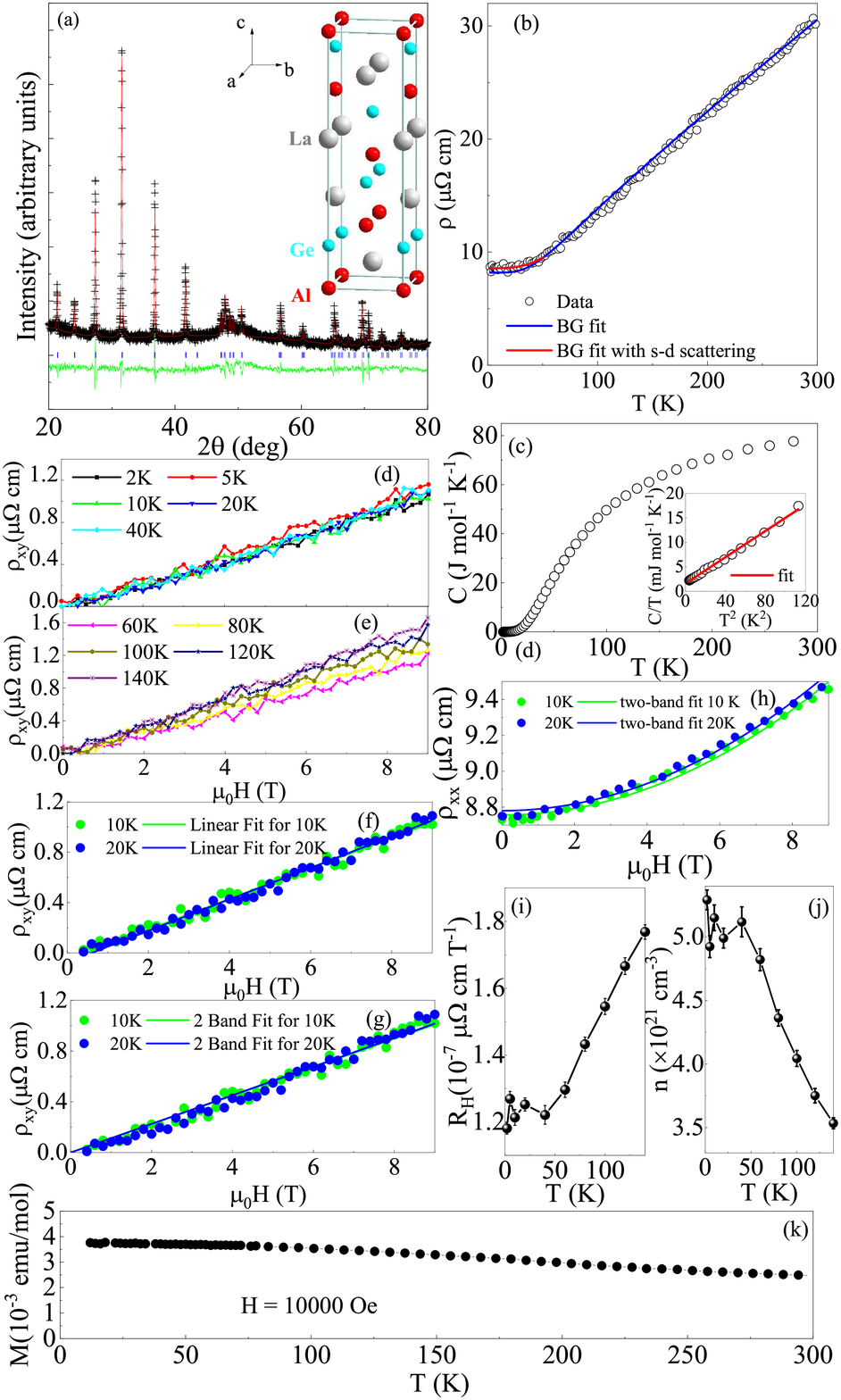}}
\caption{(Color online). (a) Powder XRD pattern of LaAlGe at room temperature:  data (+), structural model (red solid line), difference curve (green solid line, offset for clarity). The vertical tick marks represent Bragg reflections of the $I4_1/amd$ space group. Inset shows crystal structure of LaAlGe. (b) Resistivity $\rho(T)$ of LaAlGe; blue and red lines shows fitting using Bloch-Gr\"uneisen (BG) model (see text). (c) Heat capacity of LaAlGe with Debye model fit (inset). (d,e) Hall resistivity $\rho_{xy}$ measured perpendicular to the current and field out of plane and at different temperatures. Single-band (f) fits of the Hall data. Two-band fits of the magnetic-field dependent $\rho_{xy}$ (g) and resistivity $\rho_{xx}$ (h) at 10 K and 20 K. Hall coefficient $R_H$ (i), and carrier density (j) within the single-band approximation (see text). (k) Magnetization of LaAlGe in 10 kOe magnetic field.}
\label{res}
\end{figure}

Powder XRD confirms phase purity [Fig. 1(a)]. The unit cell [Fig. 1(a) inset] can be indexed into $I4_1md$ (109) space group with lattice parameters $a$ = $b$ = 0.4339(3) nm, $c$ = 1.4821(2) nm in agreement with  reported values.\cite{ZhaoJ} The crystal structure is derived from the $\alpha$-ThSi$_2$ $I4_1/amd$ prototype where La replaces Th and Ge, Al take the place of Si.\cite{kneidinger2013synthesis}

The electric resistivity [Fig.1(b)] is metallic and, in the absence of magnetic field, it is explained well by the dominant phonon scattering via Bloch-Gr\"uneisen formula:
\[\rho(T) = \rho_0+A\left(\dfrac{T}{\theta_D}\right)^n\int_{0}^{\dfrac{\theta_D}{T}}{\dfrac{z^5}{(e^z-1)(1-e^{-z})}dz},\]
where $\theta_D$ = 245(8) K is the Debye Temperature, $\rho$$_0$ = 8.67(5) $\mu$$\Omega$ cm is residual resistivity. Whereas LaAlGe resistivity is well explained by dominant phonon scattering down to 50 K with $n$ = 5, at lower temperatures $n$ = 3 provides better fit. In the low temperature limit, Bloch-Gr\"uneisen formula could be approximated with $\rho = \rho_0+AT^n$. Fit to our data [Fig. 1(b) red curve] gives  $\rho_0 = 8.58(5) \mu\Omega\times$cm and $n$ = 3.2(1). $n \sim 3$ indicates the Bloch-Wilson limit, where $n = 3$ due to $s$-$d$ interband scattering. RRR = 3.4 confirms somewhat higher degree of crystalline order when compared to thin films.\cite{Bhattarai} Heat capacity $C(T)$ is shown in Fig. 1(c). The fitting of the low-temperature data using $C$=$\gamma$$T$+$\beta$$T^3$ gives $\beta$ = 0.134(2) mJ mole$^{-1}$ K$^{-4}$ and $\gamma$ = 1.6(1) mJ mole$^{-1}$K$^{-2}$. The Debye temperature $\Theta$$_{D}$ = 243(1) K is obtained from $\Theta_{D}=(12\pi^{4}NR/5\beta)^{1/3}$ where $N$ is the atomic number in the chemical formula and $R$ is the universal gas constant. The obtained Debye temperature shows excellent agreement with Bloch-Gr\"uneisen fit results, indicating dominant phonon scattering of $\rho(T)$ where most phonon modes take part.

Hall resistivity $\rho_{xy}$ [Fig. 1(d,e)] is nearly linear in magnetic field suggesting that electronic bands at the Fermi surface are probably compensated with similar overall electron and hole concentration and with similar mobilities.\cite{ChangG,XuSY,WangQi} In the simplest approximation temperature-dependent Hall coefficient $R_H$ could be estimated from linear data fits $\rho_{xy}$ vs. $H$, as shown in [Fig.1(f)]. On the other hand, it is likely that LaAlGe features multiband electronic transport, as expected for Weyl semimetals:\cite{ChangG,XuSY,WangAFWP2}
\[\rho_{xx} = \dfrac{(n_e\mu_e+n_h\mu_h)+\mu_e\mu_h(n_e\mu_h+n_h\mu_e)B^2}{(n_e\mu_e+n_h\mu_h)^2+\mu_e^2\mu_h^2(n_h-n_e)^2B^2}\times \dfrac{1}{e},\]
\[\rho_{xy} = \dfrac{(n_h\mu_h^2-n_e\mu_e^2)+\mu_e^2\mu_h^2(n_h-n_e)B^2}{(n_e\mu_e+n_h\mu_h)^2+\mu_e^2\mu_h^2(n_h-n_e)^2B^2}\times \dfrac{B}{e},\]
where $n_e, n_h, \mu_e, \mu_h$ are carrier densities and mobilities of electron and hole-type carriers. Electron-hole compensation $n_e = n_h = n$ further simplifies to:
\[\rho_{xx} = \dfrac{1}{ne(\mu_e+\mu_h)}+\dfrac{\mu_e\mu_h}{ne(\mu_e+\mu_h)}B^2,\]
\[\rho_{xy} = \dfrac{\mu_h-\mu_e}{ne(\mu_h+\mu_e)}B.\]
By fitting magnetic-field dependent $\rho_{xy}$ [Fig. 1(g)] and $\rho_{xx}$ [Fig. 1(h)] we obtain $\mu_e$ = 262(1) cm$^2$V$^{-1}$s$^{-1}$, $\mu_h$ = 400(7) cm$^2$V$^{-1}$s$^{-1}$, $n_e$ = 1.08(1)$\times$10$^{21}$ cm$^{-3}$ and $n_h$ = 1.08(2)$\times$10$^{21}$ cm$^{-3}$ at 10K. Likewise, we obtain $\mu_e$ = 265(1) cm$^2$V$^{-1}$s$^{-1}$, $\mu_h$ = 398(8) cm$^2$V$^{-1}$s$^{-1}$, $n_e$ = 1.06(1)$\times$10$^{21}$ cm$^{-3}$ and $n_h$ = 1.06(2)$\times$10$^{21}$ cm$^{-3}$ at 20K. Therefore due to compensated Fermi surface pockets, temperature-dependent Hall coefficient and carrier density $n$ could be approximated with the single band model [Fig. 1(i,j)] where $e$ is electronic charge, $R_H = \dfrac{1}{ne}$ and $R_H$ is the slope of $\rho_{xy}$ [Fig. 1(d,e)]. Carrier concentration shows reduction above 40 K and is consistent with result obtained in thin films.\cite{Bhattarai} This is coincident with increased resistivity [Fig. 1(b)] above 40 K, indicating that $\rho(T)$ is also influenced by changes in $n$, in addition to phonon-related scattering induced by increased vibrations. Magnetization of LaAlGe is paramagnetic, somewhat enhanced at low temperatures [Fig. 1(k)].

\begin{figure}[!htb]
\centerline{\includegraphics[scale=0.35]{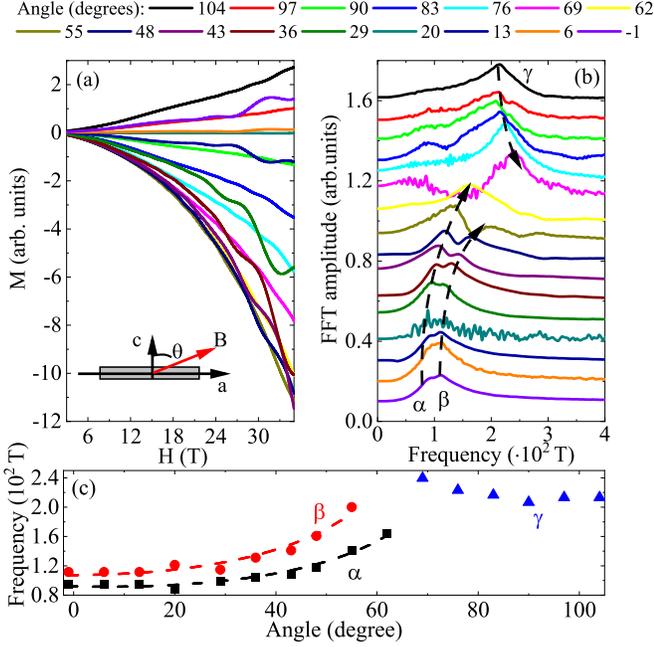}}
\caption{(Color online). (a) Angle-dependent dHvA oscillations. LaAlGe crystal installed on cantilever is rotated from -14$^\circ$ to 91$^\circ$ with data collected approximately every 7$^\circ$; (b) Fourier transform spectrum of oscillations at different angles (c) Traces of frequencies $\alpha$ and $\beta$ as a function of the angle are well fitted to quasi-2D model $\sim$1/cos$\theta$.}
\label{magnetism}
\end{figure}

Angle-dependent de Haas-van Alphen oscillations [Fig.2(a)] appear at all angles above 8 T. Figure 2(b) plots the angular evolution of Fast Fourier Transform (FFT) frequencies. Clear traces of three frequencies $\alpha$, $\beta$ and $\gamma$ are plotted individually in Fig.2(b,c). Frequencies $\alpha$ and $\beta$ increase continuously above about 0$^\circ$. The $\gamma$ frequency can be traced in a relatively narrow range between 70$^{\circ}$ and 110$^{\circ}$ and can not be fitted by either ellipsoidal or two-dimensional model. On the other hand, angular-dependence of $F_{\alpha}$ and $F_{\beta}$ is well fitted by the two-dimensional model $F(\theta)$ = A/${\cos(\theta)}$ where A is a constant.\cite{KefengSr} This observation is consistent with the possible quasi-2D Fermi surface sheets.

\begin{figure}[!htb]
\centerline{\includegraphics[scale=0.35]{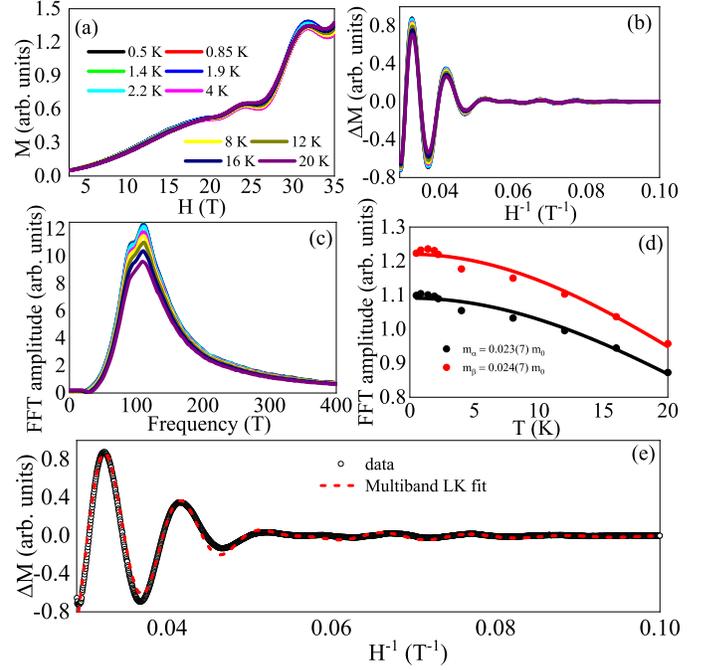}}
\caption{(Color online). Cantilever magnetization of LaAlGe at different temperatures with magnetic field applied along the $c$-axis (a) and dHvA oscillatory components obtained by smooth background subtraction (b). Fast Fourier transform spectra for dHvA (c). FFT amplitude vs temperature; LK fits are shown with the solid lines (see text).(e) Oscillatory magnetization at 0.5 K solid lines are fit of The red solid line represents the multiband LK fit.}
\label{res}
\end{figure}

Temperature dependent dHvA oscillations are also observed above 10 T in all magnetic fields [Fig. 3(a)]. Oscillation component and Fourier transform of dHvA oscillations [Fig. 3(b,c)] reveal clear signature of two frequencies, F$_{\alpha}$ = 91.9(1) T and F$_{\beta}$ = 108.3(1) T consistent with Fig. 2. This shows that multiple but closed Fermi surfaces exist in the $ab$-plane whose relation with frequencies is given by Onsager formula $F = (\hbar/2\pi e)A$ where $A$ is the cross-section area of Fermi surface perpendicular to the field: $A_{\alpha}$ = 0.87(1) nm$^{-2}$ and $A_{\beta}$ = 1.03(1) nm$^{-2}$.

The oscillatory component of electronic system with Dirac points in magnetic field is described by the Lifshitz-Kosevich formula with the Berry phase:\cite{Shoenberg}

\[\Delta M \propto -B^{\lambda} R_T R_D R_S \sin{2\pi[(F/B)-\gamma-\delta]}, \]
The $R_T$=$\alpha$$m^*$$T$/$H$$\sinh$($\alpha$$m^*$$T$/$H$), $R_D$=exp(-$\alpha$$m^*$$T_D$/$H$) and $R_S$=cos($\pi$$g$$m^*$/2) where $m^* = m/m_e$ is the effective mass of the cyclotron orbit, $T_D$ is the Dingle temperature and $\alpha = 2\pi^2k_B/e\hbar$ $\approx$ 14.69 T/K. The exponent $\lambda$ = 0 for 2D and 1/2 for 3D Fermi surface sheets. The argument of the oscillatory sine function contains phase factor (-$\gamma$-$\delta$) in which $\delta$=0 for 2D Fermi surface and $\gamma$=(1/2)-$\phi$$_B$/2$\pi$ where $\phi$$_B$ is the Berry phase. Fits of the oscillation amplitude temperature dependence to the thermal damping factor $R_T$ [Fig. 3(d)] gives effective masses $m^{*}_{\alpha}$=0.023(7)$m_0$ and $m^{*}_{\beta}$=0.024(7)$m_0$.

Since dominant Fermi surface sheets are quasi-2D, Fermi velocity $v_F$=$\hbar$$k_F$/$m^{*}$ could be estimated from the quasiparticle masses and Fermi wavevector $k_F$ which is related to the circular cross-section area of the Fermi surface as $A = \pi k_F^2$. For $F_{\alpha}$ and $F_{\beta}$ calculations give $k_{F,\alpha}$ = 0.053(1) ${\AA}$$^{-1}$, $k_{F,\beta}$ = 0.057(1) ${\AA}$$^{-1}$,  $v_{\alpha}$=2.66(1)$\times$10$^{6}$ m/s and $v_{\beta}$=2.77(1)$\times$10$^{6}$. These are large Fermi velocities, somewhat higher when compared to velocities near tilted Dirac cone-associated $W_2$ points measured in angular resolved photoemission (ARPES)\cite{XuSY} but on the other hand they are comparable to SrMnBi$_{2}$ and canonical Weyl semimetals WTe$_{2}$ or NbP (Table I).\cite{KefengSr,PengL,WangZ}

\begin{table}[h!]
\centering
\begin{tabular}{|c|c|c|}
 \hline
 Weyl Semimental&Effective Mass&Velocity\\
 \hline
 LaAlGe $\alpha$& 0.023$m_0$    &2.66$\times$10$^{6}$ m/s\\
 LaAlGe $\beta$&   0.024$m_0$  & 2.77$\times$10$^{6}$ m/s\\
 \hline
 SrMnBi$_2$ &0.29$m_0$&5.13$\times$10$^{5}$ m/s\cite{KefengSr}\\
 \hline
 WTe$_2$ $\alpha$&0.412$m_0$& \multirow{2}{*}{3.09$\times$10$^{5}$ m/s\cite{PengL}}\\
 WTe$_2$ $\beta$&0.482$m_0$&\\
 \hline
 NbP & 0.1$m_0$ & 1.8$\times$10$^{5}$ m/s\cite{WangZ}\\
 \hline
\end{tabular}
\caption{Comparison of Weyl semimetal LaAlGe, SrMnBi$_2$, WTe$_2$, NbP}
\label{table:1}
\end{table}

Multiple frequency oscillations are well treated by the linear superposition of several single-frequency oscillations using multiband LK formula for data taken at 0.5 K.\cite{JHu1,JHu2,JHu3,XuS} The fit [Fig. 3(e)] yields Berry phase for different Fermi surface sheets $\phi$$_{B,\alpha}$ = 0.88(1)$\pi$ for F$_{\alpha}$ and $\phi$$_{B,\beta}$ = 0.96(1)$\pi$ for F$_{\beta}$. Topologically non-trivial Berry phases confirm the linear dispersion of Weyl fermions for the observed Fermi surface pockets.

The fitted Dingle temperatures are $T_{D,\alpha}$ = 159(4) K and $T_{D,\beta}$ = 336(4) K. From $T_{D} = \frac{\hbar}{2\pi{k_B}\tau_Q}$ we evaluate scattering rate $\tau$$_{\alpha}$ = 7.64(1)$\times$10$^{-15}$ s and $\tau$$_{\beta}$ = 3.62(1)$\times$10$^{-15}$. Mobility estimate using $\mu$ = $e\tau/m^*$ gives $\mu$$_{\alpha}$ = 584(1) cm$^2$V$^{-1}$s$^{-1}$ and $\mu$$_{\beta}$ = 265(1) cm$^2$V$^{-1}$s$^{-1}$; when compared to mobility estimate from Hall effect at 10 K (Fig. 1) this could imply somewhat higher temperature dependence of the hole pocket.

Very small effective masses are consistent with Weyl points in LaAlGe Fermi surface. Since other Weyl points are 60 meV to 130 meV away from the Fermi level, the quasi-2D Fermi surface parts that correspond to observed frequencies F$_{\alpha}$, F$_{\beta}$ but also F$_{\gamma}$ should correspond to some of the type-II $W_2$ Weyl cones with tilted dispersion away from the $k_z$ plane.\cite{XuSY}

We estimate electronic specific heat from the effective mass obtained in quantum oscillation experiment using quasi-two-dimensional approximation:\cite{Ca3Ru2O7}
\[\gamma_{N} = \sum_{i}\frac{\pi N_{A}k_{B}^{2}ab}{3\hbar^{2}}m^*\]
where $a$ = $b$ = 0.4339(2) are the tetragonal plane lattice parameters, $m^*$ is the quasiparticle mass, $N_A$ is Avogadro's number, $k_B$ is Boltzmann's constant and $\hbar$ is Planck's constant. From the sum over all Fermi pockets which include four $W_2$s on the top of Brillouin zone for $\alpha$ and four on top for $\beta$ frequencies, four $W_2$s on the bottom for $\alpha$ and four for $\beta$,  $\gamma_{N}$ = 0.70(5)(1) mJ mol$^{-1}$ K$^{-2}$ can be obtained.\cite{XuSY} This is less than the value obtained from fits in [Fig. 1(c)] and indicates that not all portions of Fermi surface detected in quantum oscillations contribute to electronic specific heat.

In summary we showed that LaAlGe features electronic transport dominated by Weyl nodes with very small effective masses and high Fermi velocities. The dominant type-II Weyl cones at the Fermi surface are possibly quasi-2D, despite the absence of two-dimensional crystallographic units as in materials with van der Waals bonds. Such conducting states coupled with magnetic moments are stable in the presence of long-range Coulomb interactions,\cite{ChangG,LeeY2} which points to possibility of robust spin polarized currents in future devices.

Work at Brookhaven National Laboratory was supported by US DOE, Office of Science, Office of Basic Energy Sciences under contract DE-SC0012704. A portion of this work was performed at the National High Magnetic Field Laboratory, which is supported by the National Science Foundation Cooperative Agreement No. DMR-1644779 and the State of Florida.

$^{\P}$ Present address: Los Alamos National Laboratory, MS K764, Los Alamos NM 87545
$^{\ddag}$ petrovic@bnl.gov

\section{DATA AVAILABILITY}
The data that support the findings of this study are available from the corresponding author upon reasonable request.

\end{document}